\documentstyle[twocolumn,aps,prl,epsf,floats]{revtex}

\begin{document}

\title {Electronic Raman Scattering in  Superconducting Cuprates}
\author{Andrey V. Chubukov$^{1}$,  Dirk K. Morr$^{2}$, and
Girsh Blumberg$^{3}$}
\address{$^{1}$
Department of Physics, University of Wisconsin, Madison, WI 53706}
\address{$^{2}$
NSF Science and Technology Center for Superconductivity and\\
Loomis Laboratory of Physics, University of Illinois at Urbana-Champaign,
1110 W. Green St., Urbana, IL 61801}
\address{$^{3}$
Bell Laboratories, Lucent Technologies, 700 Mountain Ave., Murray Hill, NJ
07974}
\date{received July 22 by M.Cardona}
\draft

\maketitle

\begin{abstract}
We show that the novel features observed in
Raman experiments on optimally doped and underdoped
Bi-$2212$ compounds in $B_{1g}$ geometry can be
explained by a strong fermionic self-energy due to the
interaction with spin fluctuations.
We compute the Raman intensity $R(\omega)$ both above and below $T_c$, and
show that in both cases $R(\omega)$ progressively deviates,
with decreasing doping, from that in a
Fermi-gas due to increasing contribution
from the fermionic self-energy. We also show that
the final state interaction increases with decreasing doping and gradually
transforms the $2\Delta$ peak
in the superconducting state into a pseudo resonance mode below $2\Delta$.
We argue that these results  agree well with the
experimental data for Bi-$2212$.
\end{abstract}
\pacs{keywords: A Superconductors, E Inelastic Light Scattering, D
Electron Electron Interactions.}

\narrowtext

\begin{figure} [t]
\begin{center}
\leavevmode
\epsfxsize=7.5cm
\epsffile{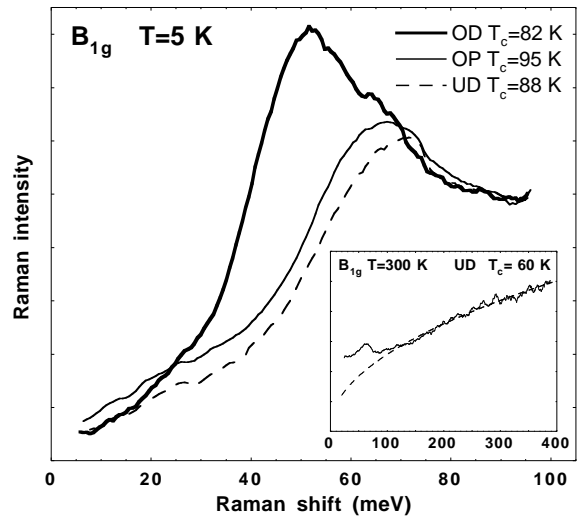}
\end{center}
\caption{
The $B_{1g}$ electronic Raman scattering spectra at $T = 4$~K for overdoped
(OD), optimally  doped (OP) and underdoped (UD)
Bi$_{2}$Sr$_{2}$CaCu$_{2}$O$_{8 \pm \delta}$ (Ref.~\protect\cite{Blu98}).
The inset shows  the high energy continuum in the normal
state for the underdoped $T_c = 60$~K sample and its fit to
the $\omega^{1/2}$ behavior.}
\label{exp}
\end{figure}

The form of the fermionic spectral function in optimally doped and
underdoped cuprate superconductors has been the
subject of intensive experimental and theoretical studies over the last few
years.
ARPES and neutron experiments demonstrated
that  the fermionic spectral function
undergoes a substantial evolution with decreasing doping,
and for underdoped cuprates is very different from
the one in a Fermi-gas (FG)~\cite{photodata}.
Raman scattering in $B_{1g}$
geometry is another spectroscopy to study
the electronic properties in this momentum region.
For electronic Raman scattering, the intensity
$R(\omega)$ is in general given by
the imaginary part of the fermionic particle-hole bubble at small external
momentum ${\bf q}$ and finite frequency, weighted with the Raman
vertices~\cite{Abrikosov,Klein84,Shastry,Devereaux}.
In this paper we focus on the $B_{1g}$ Raman scattering.
In $B_{1g}$ geometry, the Raman vertex
$V_{B_{1g}} \propto \cos k_x - \cos k_y$,
and the scattering thus mostly probes the vicinity of
$(0,\pi)$~\cite{Devereaux}.
Recent Raman experiments
on overdoped, optimally doped and underdoped Bi$-$2212  demonstrated
that the $B_{1g}$ Raman intensity $R(\omega)$ undergoes
significant changes with decreasing
doping~\cite{Kendziora,Hackl96,Blu98,Quilty}, and progressively
deviates from predictions of a FG theory.
Indeed, according to the FG theory,
$R(\omega)$ in the normal state is finite only
for nonzero external momentum $q$, and vanishes
at $\omega >  v_F q$, where $v_F$ is the Fermi velocity~\cite{Platzman}.
In the superconducting state,
$R(\omega)$ is finite due to a particle-hole mixing
and possesses a peak at $\omega = 2\Delta$
where $\Delta$ is the maximum of the superconducting gap
$\Delta_k$~\cite{Klein84}.
Below $2\Delta$, the intensity scales as $\omega^3$ at
small frequencies due to  the presence of  nodes
in the $d-$wave gap~\cite{Devereaux}.
In Fig.~\ref{exp}
 we present the experimental Raman intensity
for several Bi$-$2212 compounds both below and above $T_c$ \cite{Blu98}.
In the overdoped, $T_c =82K$ material,
the behavior of $R(\omega)$  is qualitatively
consistent with the FG theory, i.e., it is featureless above $T_c$ (not shown),
while at $T< T_c$,
$R(\omega) \propto \omega^3$ at small
frequencies~\cite{Staufer}, and exhibits a sharp peak at $2\Delta \approx
50$ meV. At and below optimal doping,  however, the form of $R(\omega)$
is  inconsistent with the FG theory.
In the normal state, the intensity
 increases  with frequency as $R(\omega) \sim \sqrt{\omega}$,
 and saturates at a few hundred meV~\cite{Hackl96,Blu98,Staufer}.
(see the inset in Fig.~\ref{exp}). In the superconducting state,
the key experimental observation is that with underdoping,
the peak in $R(\omega)$ occurs at progressively
smaller frequency than
the "$2\Delta$" extracted from
tunneling experiments~\cite{tunn}, and
almost saturates at about $75$ meV~\cite{Blu98}.
Simultaneously, the low frequency behavior becomes predominantly
linear in $\omega$ with
a kink around $40-50$ meV, while above the peak
 the intensity develops a dip at
about $90$ meV and at even larger frequencies recovers the normal
 state $\sqrt{\omega}$ form.

In this Letter, we make two points. First, we argue
 that the inconsistency of
the experimental results for Bi-$2212$ at and below optimal doping
 with the FG scenario
 indicates that the fermionic self-energy is large and substantially
modifies the form of the fermionic Green's function both in the
normal and in the superconducting state.
Phenomenologically, this effect
was considered in~\cite{Varma89}. Here we
perform a microscopic analysis within a spin-fermion model
in which the fermionic self-energy emerges
due to an interaction with overdamped spin fluctuations and
is the largest for fermions in the
vicinity of $(0,\pi)$.
Since the Raman intensity in $B_{1g}$ geometry predominantly probes the
region near $(0,\pi)$, it directly reflects the changes in the
fermionic spectrum imposed by the large self-energy.
Second, we show that  a
magnetically induced final state interaction
replaces the $2\Delta$ peak in $R(\omega)$
by a pseudo resonance peak at a smaller frequency $\omega_{res}$.
Near optimal doping, this effect is almost unobservable, but
it becomes visible in underdoped cuprates and explains the discrepancy between
the Raman peak frequency and the one in the density of states.
\begin{figure} [t]
\begin{center}
\leavevmode
\epsfxsize=7.5cm
\epsffile{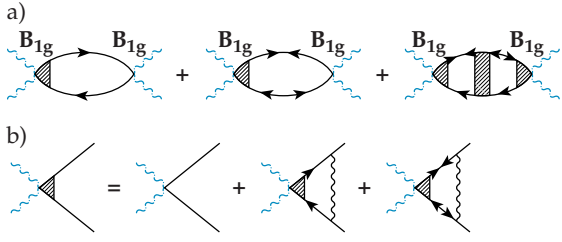}
\end{center}
\caption{The full $B_{1g}$
Raman intensity in a superconductor (a), and the full
$B_{1g}$ Raman vertex (b). The wavy line and dashed rectangle
are the fully renormalized effective
interaction in  $d-$wave and $s-$wave channels, respectively.}
\label{fig3}
\end{figure}

We now turn to the calculations. The
 Raman intensity in a superconductor is
given by a set of
fermionic bubbles made of  normal and anomalous Green's
functions~\cite{Schrieffer,m-z}
(Fig.\ref{fig3}).
We first compute $R(\omega)$ without final
state interaction.
 In this approximation, only the first two diagrams in
Fig.~\ref{fig3}a contribute (with bare vertices), and we have
\begin{eqnarray}
&&R(\omega) \propto Im \int d k~ d \Omega ~
~V^2_{B_{1g}}({\bf k})\nonumber \\
&&\times (G_{sc} (k,\Omega_+) ~
G_{sc} (k,\Omega_-) +
F(k,\Omega_+)~ F(k, \Omega_-)),
\label{R_sup}
\end{eqnarray}
where $\Omega_\pm = \Omega \pm \omega/2$ and
\begin{eqnarray}
G_{sc} (k, \omega) &=& G^{-1}_n (-k, -\omega)/
(G^{-1}_n (k, \omega)~G^{-1}_n (-k, -\omega) + \Delta^2_k)
\nonumber \\
F (k, \omega) &=&
i \Delta_k/(G^{-1}_n (k, \omega)~G^{-1}_n (-k, -\omega) +
\Delta^2_k) \ .
\end{eqnarray}
Here $\Delta_k$ is the $d$-wave superconducting gap, and
$G^{-1}_n (k,\omega) =\omega - \epsilon_k + {\bar g}^2
\Sigma({\bf k},\omega)$,  where
${\bar g}$ is a dimensionless spin-fermion coupling, and
$\Sigma ({\bf k},\omega)$ is the $\Delta$-dependent
fermionic self-energy. Theoretically,
${\bar g}^2 \propto \xi^{-1}$ where $\xi$ is the magnetic
correlation length~\cite{chubukov}.
In strongly overdoped cuprates,
 ${\bar g} \leq 1$, and the system behavior
resembles that in a FG.  However, at and below optimal doping,
${\bar g} \geq 1$ in which case the
self-energy overshadows the bare $\omega$ term in the Green's function, i.e.,
$G^{-1}_n (k,\omega) \approx  {\bar g}^2 \Sigma(k, \omega)
 -\epsilon_k$.

The form of the fermionic self-energy
is an input for our Raman calculations. Obviously, the self-energy is the
largest near $(0,\pi)$ and symmetry related points where the scattering by
nearly antiferromagnetic spin fluctuations is the strongest. The self-energy in
this $k-$range also strongly depends on doping as evidenced by ARPES data.
Since our goal is to relate the doping dependent changes in $R(\omega)$ with
those in $\Sigma (k, \omega)$, we restrict our consideration to the vicinity of
$(0,\pi)$ where $\Delta_k$ is close to its maximum value $\Delta$.
We will argue below that for strong
coupling, the $B_{1g}$ Raman intensity is dominated by $k$ near $(0,\pi)$
down to $\omega \ll \Delta$, and crosses over to $\omega^3$ behavior due to
the nodes of the d-wave gap only at vanishingly small frequencies.

It has been argued~\cite{chubukov,chu-morr} that for ${\bar g} \gg 1$,
$\Sigma (k,\omega)$ near $(0,\pi)$ is
independent of the quasiparticle energy up to corrections
$O((\log {\bar g})/{\bar g}^2)$, and in a superconductor
behaves as $\Sigma (\omega) \propto
 \omega$ at
$\omega \ll 2{\bar \Delta}$, and
as  $\Sigma(\omega) \propto e^{i\pi/4} \sqrt{|\omega|} sgn \omega$
at $\omega \gg max (2 {\bar \Delta},\omega_{sf})$, where
 ${\bar \Delta}= \Delta/{\bar g}$ is a measured gap, and
$\omega_{sf} \propto \xi^{-2}$ is a typical relaxation frequency of spin
fluctuations.
The physical reasoning here is twofold. First,
in the normal state, the
scattering by nearly-critical overdamped spin fluctuations
 yields
$\Sigma(\omega) = 2\omega/(1 + \sqrt{1 - i |\omega|/\omega_{sf}})$
which displays a crossover
 from a Fermi-liquid behavior at $\omega <
\omega_{sf}$ to a quantum-critical behavior for $\omega > \omega_{sf}$
where $\Sigma(\omega) \propto
\sqrt{\omega}$~\cite{chubukov}.
Second, in a superconductor, the strength of the
scattering by spin fluctuations
is reduced below $2{\bar \Delta}$ due to a feedback effect on the spin
damping~\cite{chu-morr}.
 The  calculation of the full $\Sigma (\omega)$  is rather
involved and requires one to solve a set of two coupled complex
integral equations for
the fermionic self-energy and the spin polarization operator.
Below we use the
approximate
 self-consistent solution of this set of equations~\cite{chu-morr}
which
correctly reproduces the behavior of
$\Sigma (\omega)$ at  large and small frequencies. In the
latter case it yields $\Sigma (\omega) \approx \omega$, i.e.,
the peak in the spectral function occurs right at $\omega = {\bar
\Delta}$.

We now proceed with $R(\omega)$.
We assume that
the density of states near $(0,\pi)$ depends only weakly on $\epsilon_k$
 and replace the $k-$integration in
(\ref{R_sup}) by the integration over $\epsilon_k$.  We then obtain
(neglecting overall prefactor)
$R(\omega) = Im \chi (\omega)$ where
\begin{equation}
\chi (\omega) = -\int_0^{\infty} d\Omega
\frac{{\bar \Delta}^2 -\Sigma(\Omega_+)\Sigma(\Omega_-) +
D(\Omega_+)D(\Omega_-)}{D(\Omega_+)D(\Omega_-)(D(\Omega_+) +
D(\Omega_-))}~ +~C,
\label{ram}
\end{equation}
and $D (\Omega_\pm) = \sqrt{\Sigma^2 (\Omega_\pm)-{\bar \Delta}^2}$.

The constant $C>0$ is a real number. Its presence in (\ref{ram}) 
is related to the fact
that Eq. (\ref{R_sup}) with
$k-$integration substituted by the integration over 
$\epsilon_k$ lacks convergence.
In this case, one cannot simply
interchange  frequency and energy integrations and has to include a
regularization procedure~\cite{vr}.
The value of $C$ is
irrelevant for the calculations of the Raman intensity without vertex
corrections as it does not contribute to $Im \chi (\omega)$.
It however becomes
relevant when one includes the effects of the final state interaction which
accounts for the renormalization of the Raman vertex (see
below). These  vertex
corrections involve the $d-$wave component of
the effective interaction, $\Gamma_d$ which
decreases at large frequencies and therefore provides the physical
regularization of the Raman bubble.
In the spin-fluctuation approach,
$\Gamma_d$ starts decreasing at $\omega \sim
\epsilon^2_F/{\bar g}$~\cite{chubukov}.
Another regularization is provided by the fact
that the integral over $\epsilon_k$ is cut at $\epsilon_F$. Obviously,
 the value of
$C$ depends on the ratio $\epsilon_F/{\bar g}$.
For $\epsilon_F \gg {\bar g}$ which is the case at weak coupling,
the cutoff in a momentum space dominates.
In this situation, an adequate way to evaluate the Raman bubble is to
integrate first over frequency. This is how previous calculations of
$R(\omega)$ have been performed~\cite{Klein84,Devereaux}. One then obtains
$C = 1$. However, in the strong coupling limit
$\epsilon_F \ll {\bar g}$, which is likely to be satisfied in underdoped
cuprates, a more adequate way to
 evaluate the Raman bubble is to integrate first over $\epsilon_k$ as we
did. In this case, $C=0$.

We now analyze Eq. (\ref{ram}) first in the normal state and then in the
superconducting state.
In the normal state, a substitution of the explicit form of
$\Sigma ({\omega})$ into
(\ref{ram}) yields $R(\omega) = R(\omega_{sf})~
\Phi (\omega/\omega_{sf})$
where $\Phi (x) \approx 1.07 x$ for $x \ll 1$, and $\Phi(x) \approx 1.73
\sqrt{x}$ for $ 1 \ll x \ll {\bar g}^4$. We see that at
 $\omega > \omega_{sf}$, relevant to experiments,
$R(\omega) \propto \sqrt {\omega}$.
We fitted the normal state data by this form and found almost perfect
agreement with the experiment (see the inset of Fig.~\ref{exp}).
 For even larger $x \geq {\bar g}^4$, the bare
$\omega$ term in the quasiparticle Green's function begins to dominate,
 and the theoretical $R(\omega)$ saturates,
passes through a very broad maximum at $x = 2.24 {\bar
g}^4$ and then slowly decays as $1/\sqrt{x}$. This also agrees with the data.
A similar behavior was
 found numerically in Ref.~\cite{DevKampf}.

\begin{figure} [t]
\begin{center}
\leavevmode
\epsfxsize=6.5cm
\epsffile{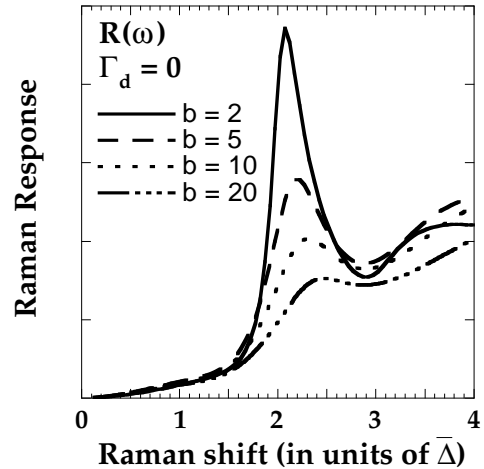}
\end{center}
\caption
{A calculated Raman intensity,
Eqs. (\protect\ref{ram},\protect\ref{res})
for different values of $b = {\bar \Delta}/\omega_{sf}$. Larger $b$
correspond to stronger self-energy corrections which destroy fermionic
coherence.
Observe the broadening of the peak with increasing $b$ and the development
of the dip at around $3\bar \Delta$.}
\label{numram}
\end{figure}
In the superconducting state, the form of $R(\omega)$ clearly
depends on the ratio
 $b={\bar \Delta}/\omega_{sf}$.
Experimentally, $b \leq 1$ in
strongly overdoped cuprates, and $b \gg 1$ in strongly underdoped
cuprates~\cite{photodata}.
For small $b$,
Eq.(\ref{ram}) expectedly reproduces the
FG result for the Raman intensity: $R(\omega) =0$ for $\omega < 2{\bar \Delta}$
and $R(\omega) \propto (\omega \sqrt{\omega^2 -
 4{\bar \Delta}^2})^{-1}$ for
$\omega > 2 {\bar \Delta}$. The inclusion of the $k-$ dependence of the gap
corrects this behavior at small
frequencies where it yields a finite
$R(\omega) \propto (\omega/{\bar \Delta})^3$~\cite{Devereaux},
and also very near
$2{\bar \Delta}$ where it changes the square root
singularity to a logarithmic one.
For large $b$, however, the form of the Raman intensity substantially
deviates from the FG result.
We computed numerically  the Raman intensity for different
values of $b$, and present the results in Fig.~\ref{numram}.
We emphasize four key features in $R(\omega)$:
(i) the $2{\bar \Delta}$ peak is still present even for large $b$,
(ii) with increasing $b$, the peak frequency becomes progressively larger than
$2{\bar \Delta}$,
(iii) there is a dip in $R(\omega)$ above the peak,
(iv) at small frequencies, $\omega \ll 2 {\bar \Delta}$, $R(\omega)$ is
predominantly linear in $\omega$.
These features in the Raman intensity can all be understood analytically
by analyzing Eq. (\ref{ram}) in various limits.

The broadening of the peak, the dip and the linear behavior
at low frequencies all agree with the features present in the data in
Fig.~\ref{exp}. However, one key disagreement with the data remains. Namely,
 we  calculated the density of states
$N(\omega) = \int dk Im G_{sc} (k,\omega)$
along the same lines as the Raman intensity
and found that the peak in $N(\omega)$ is located
at almost  exactly a half
of the peak frequency in $R(\omega)$, although both peak frequencies shift to
larger values with increasing $b$. The data, we remind, show that
with underdoping, the peak  in $R(\omega)$  occurs at progressively lower
frequencies than twice the peak frequency in $N(\omega)$.

We now argue that the experimentally observed downturn renormalization of the
peak in $R(\omega)$ compared to twice that in $N(\omega)$ is due to a fact that
the
final state interaction between scattered quasiparticles gives rise to
a bound state below $2{\bar \Delta}$.

The
final state interaction in both $s-$wave and $d$-wave superconductors
have been considered several times in
the literature~\cite{Klein84,Schrieffer,m-z,devvert}.
It gives rise to corrections
to the  particle-hole vertex (Fig.~\ref{fig3}b), and also
introduces an additional scattering process
which mixes
particle-hole and particle-particle channels~\cite{Klein84,m-z}
(Fig.~\ref{fig3}a).
A simple experimentation shows that
the vertex corrections to the
$B_{1g}$ vertex involve the $d-$wave component
of the effective interaction in the zero-sound channel, while the "mixed"
diagram (the last one in Fig.~\ref{fig3}a) contains the fully renormalized
$s-$wave vertex in the particle-particle channel.

Several authors argued~\cite{devvert}
 that for a $d-$wave superconductor with
spin-independent interaction, the effective $d-$wave coupling in the zero-sound
channel is repulsive, and the bound state
does not appear unless the mixed diagram prevails over the 
conventional, RPA-type
vertex renormalization.
We, however, demonstrate below that for magnetically-mediated $d-$wave
superconductivity
there is an additional sign change between vertices in the zero-sound
and the Cooper channels, and this eventually gives rise to attractive
$d-$wave coupling in the zero-sound channel.

Indeed, consider the
effective interaction between fermions mediated by the exchange of spin
fluctuations with momenta near $Q$. We have
$\Gamma = - g^2~ \chi ({\bf q},\Omega)~
{\bf \sigma}_{\alpha \beta} {\bf \sigma}_{\gamma
\delta}$
where ${\bf q} = {\bf k}- {\bf k}^\prime$ and $\Omega = \omega - \omega^\prime$
are transferred momentum and frequency, respectively.
To simplify the discussion on the sign of the interaction,  assume that
$\Gamma$
has a dominant $d-$wave partial amplitude $\Gamma_d$, i.e.,
\begin{equation}
\Gamma (k - k^\prime, \omega - \omega^\prime)
\approx d_k~d_{k^\prime}~\Gamma_d~
{\bf \sigma}_{\alpha \beta} {\bf \sigma}_{\gamma \delta}
\label{1}
\end{equation}
where $d_k$ are $d-$wave eigenfunctions.
In general, $\Gamma_d$ depends on the transferred
frequency. However,  one can straightforwardly demonstrate that
for a relaxational form of the spin susceptibility,
the frequency dependence of  $\Gamma_d$ becomes relevant at  frequencies
$\omega \sim \epsilon^2_F/{\bar g} \gg {\bar \Delta}$~\cite{chubukov}.
The
vertex renormalization on the other hand is
mostly determined by  much smaller
$\omega \sim {\bar \Delta}$.
In this situation, $\Gamma_d$ can, to a good accuracy, be
approximated by a constant.

Let us first verify that there is a superconducting instability in the
spin-singlet particle-particle channel.
Substituting (\ref{1}) into the equation for the full particle-particle
vertex and making use of the identity
\begin{equation}
{\bf \sigma}_{\alpha \beta} {\bf \sigma}_{\gamma \delta} = T -3S
\end{equation}
where $T,S = (\delta_{\alpha \beta} \delta_{\gamma \delta} \pm \delta_{\alpha
\delta} \delta_{\gamma \beta})/2$ are triplet and singlet spin configurations,
respectively, we obtain (using for simplicity
Fermi gas Green's functions)
\begin{equation}
\Gamma^{tot}_S =
-3 \Gamma_d/(1 + 3 \Gamma_d~ L)
\end{equation}
where $L = log~(\omega_{max}/T) >0$.
We see that the behavior of the total particle-particle vertex in the
 spin singlet channel
depends on the sign of $\Gamma_d$. 
Several authors have demonstrated that near the antiferromagnetic instability,
$\Gamma_d <0$~\cite{pines}. This obviously implies the superconducting
instability. However, if the interaction were
spin independent, then the $d-$wave instability would require a {\it positive}
$\Gamma_d$.

Let us turn to the $B_{1g}$ Raman intensity.
The simplest way to check the sign of the vertex correction is to
consider  a Fermi gas in the
normal state, and compute the
density-density correlator
at zero frequency and at a finite momentum.  In the normal state,
the mixed diagram does not contribute, and the ladder series of vertex correction
diagrams can be easily summed up. We found
\begin{equation}
V^{full}_{B_{1g}} (q) = \frac{V_{B_{1g}} (q)}{1 + \Gamma_d~ \chi_d (q)}
\label{2}
\end{equation}
where $V_{B_{1g}}$ is the bare $B_{1g}$ vertex, and
\begin{equation}
\chi_d (q) = \int d{\bf k}~(d_k)^2 \frac{\Theta (\epsilon_+) -
\Theta(\epsilon_-)}{\epsilon_+ - \epsilon_-}
\label{chi_d}
\end{equation}
is the uniform $d-$wave susceptibility ($\epsilon_{\pm} = \epsilon_{k \pm
q/2}$, and
$\Theta (x) = 1$ when $x>0$ and $\Theta
(x) =0$ when $x<0$). Obviously, $\chi_d (q) >0$. We see that for negative
$\Gamma_d$, the $B_{1g}$ vertex in the particle-hole channel is enhanced, i.e.,
the effective interaction in this channel is attractive. We show below that
in the superconducting state 
this attraction gives rise to a pseudo-resonance below $2{\bar \Delta}$.
Previous analytical
studies~\cite{devvert} obtained the same expression as in
(\ref{chi_d}), but
they considered {\it spin-independent} $d-$wave interaction, and
therefore set $\Gamma_d >0$.  In this situation, vertex corrections
reduce the $B_{1g}$ vertex and do not give rise to a resonance behavior.

The conclusion that vertex corrections reduce the
 $B_{1g}$ vertex has recently been
reached in numerical studies of the Hubbard model~\cite{dahm}.
The reasons for the discrepancy with our analysis are not clear to us
because the effective interaction selected
in~\cite{dahm} is apparently also mediated by spin fluctuations.

\begin{figure} [t]
\begin{center}
\leavevmode
\epsfxsize=8.5cm
\epsffile{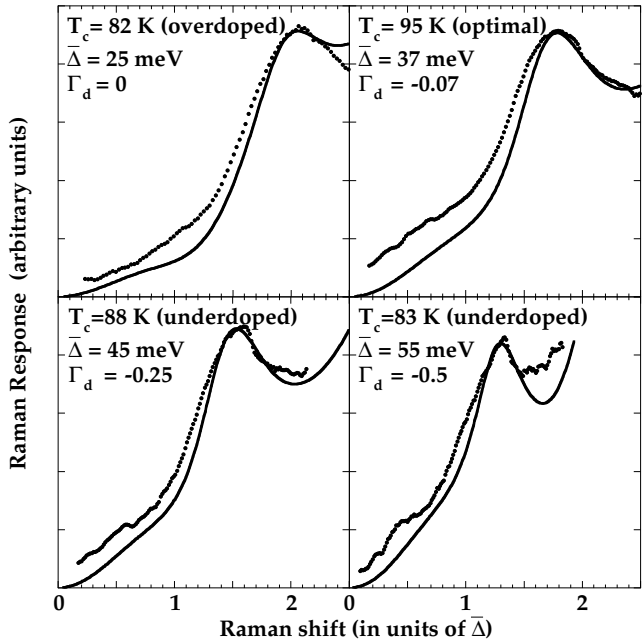}
\end{center}
\caption
{The fits of the theoretical Raman intensity
with final state interaction, $R_{full} (\omega)$, Eq.~\protect\ref{res}
to the experimental data for slightly overdoped,
optimally doped and underdoped $Bi$2212 materials.
We use the theoretical fact that in the absence of the
final state interaction, the peak frequency in
the Raman intensity, even at strong coupling, is almost exactly twice the
peak frequency in the density of states, and extract an effective
${\bar\Delta}$
from the photoemission and tunneling data~\protect\cite{photodata,tunn}.
The parameter $\Gamma_d$
measures the strength of the vertex corrections which, we argue, give rise
to a pseudo-resonance in $R_{full} (\omega)$. Observe that with underdoping,
the peak in $R_{full} (\omega)$
progressively deviates down from $2\bar \Delta$.
The presented fits are for $b=20$.
Fits using smaller $b$ reproduce the shape of $R(\omega)$ but yield sharper
peaks than in the data.}
\label{fig4}
\end{figure}

We now continue our analysis of the
superconducting state. The vertex renormalization is given by a set of
diagrams in Fig.~\ref{fig3}.
For a spin-mediated interaction,  the $s-$wave coupling
is repulsive such that the last diagram
in Fig.~\ref{fig3}a does not contain a low-energy resonance mode and can be
safely neglected. We are then left with
the ladder series of vertex correction diagrams. For a frequency independent
$\Gamma_d$, the series of vertex corrections is then
geometrical and yields
\begin{equation}
R_{full} (\omega) =
 \frac{Im \chi(\omega)}{(1 + \Gamma_d Re \chi(\omega))^2 +
(\Gamma_d Im \chi(\omega))^2}
\label{res}
\end{equation}
 where, we remind, $\Gamma_d <0$, and
$\chi(\omega)$ is given by (\ref{ram}).
Now recall that in a superconductor
$Im \chi(\omega)$ is small at $\omega < 2{\bar \Delta}$.
Evaluating  $Re \chi(\omega)$ we found that
it is {\it positive} below  $\omega < 2{\bar \Delta}$.
Thus for small frequencies, we found
$Re \chi(\omega) = A (\omega/{\bar \Delta})^2$ where in a FG,
$A=1/3$.
Substituting this result into (\ref{res}), we find that
 $R_{full}$ possesses
a resonance peak below $2{\bar \Delta}$,
at a frequency $\omega = \omega_{res}$ where
$|\Gamma_d| Re R (\omega_{res}) =1$.
Actually,  the solution for $\omega_{res}$ exists
already at weak coupling simply because in a FG, $Re \chi(\omega)$
 diverges logarithmically
as $\omega$ approaches $2{\bar \Delta}$ from
below.  As the coupling increases, the peak position progressively
deviates downwards from $2{\bar \Delta}$.
A similar reasoning has been previously applied to explain the presence of the
resonance peak in the neutron scattering data below $T_c$~\cite{neutrons}.

Indeed, the bound state which we found is
only a pseudo resonance because in a $d-$wave superconductor,
$Im \chi (\omega)$ is finite
 for all $\omega \neq 0$ because of the nodes of
the gap. On the other hand,
the very existence of the peak only requires a reduction of
$Im \chi (\omega)$ below $2{\bar \Delta}$ which
is a natural consequence of a reduction of the fermionic spectral weight
at low frequencies in a superconductor. Moreover, as the spectral weight
reduction occurs already in the pseudogap
regime~\cite{photodata}, our theory predicts
 that the Raman peak should survive above $T_c$ and disappear only
at a temperature where the pseudogap behavior becomes invisible.

In Fig.~\ref{fig4} we fitted the data from Fig. \ref{exp} to
Eq. (\ref{res}) using $\Gamma_d$ as an adjustable parameter. We
found that in overdoped cuprates, the pseudo resonance frequency is almost
indistinguishable from twice the peak frequency extracted from the
tunneling data~\cite{tunn}.
However, as the system moves towards lower doping, $|\Gamma_d|$ obviously
increases, and the peak in $R_{full}(\omega)$ moves to
progressively lower frequencies compared to twice the peak frequency of
the tunneling data.

One more point. For $s-$wave case,  weak coupling calculations~\cite{m-z}
 have shown that for an
attractive zero-sound coupling, $R_{full} (\omega)$ has two peaks, one at
$\omega_{res}$ and another near $2{\Delta}$.
We did not  find indications for a two
peak structure in $R_{full}
(\omega)$. The reason is that in our case,
the peak in $R(\omega)$ is already rather broad.
Furthermore, we found numerically
that the overall shape of $R_{full} (\omega)$ does not change much
compared to that in $R(\omega)$, i.e., the dip and the linear behavior at small
frequencies are present also in $R_{full} (\omega)$ (see Fig.~\ref{numram}).
>From this perspective, the shape of the Raman intensity is mostly determined by
the self-energy corrections, while the position of the Raman peak is determined
by the resonance in the Raman vertex.

To summarize,
we have demonstrated that the experimentally observed doping evolution
of the $B_{1g}$ Raman intensity in cuprates can be explained by an
interaction with spin fluctuations.
We argued
that for the optimally doped and underdoped materials our results capture all
salient features of the experimental data in Fig.\ref{exp}:
(i) the $\sqrt\omega$ behavior of the intensity
in the normal state,
 (ii) a predominantly linear low frequency
behavior of $R(\omega)$ in a superconductor,
(iii) a  reduction of the peak
amplitude with decreasing doping and
a development of a dip above the peak,
and (iv) a progressive downturn deviation of the Raman peak position compared
to the distance between the peaks in the tunneling density of states.
Finally, the prediction that the Raman peak  survives in the
pseudogap regime is also consistent with the
data~\cite{Blu98}.

An issue which we didn't address in this paper is the experimentally observed
strong discrepancy between the Raman data in $B_{1g}$ and $A_{1g}$
geometries~\cite{diff}.
This discrepancy (e.g.,  different locations of the peak frequencies) is still
unexplained~\cite{cardona} and clearly calls for more theoretical work on Raman
scattering.

It is our pleasure to thank T. Devereaux,
  R. Gatt, R. Joynt, M. V. Klein, A. Millis, H. Monien,
D. Pines and J. Schmalian
 for useful conversations. The research was supported by NSF grant DMR-9629839
(A.C.) and by the NSF cooperative agreement
DMR91-20000 through STCS (D.M. and G.B.).

\end{document}